\documentclass[11pt,a4paper, preprint,  superscriptaddress]{revtex4}
\usepackage[utf8]{inputenc}
\usepackage{hyperref}
\usepackage{pdflscape}
 \usepackage{amsmath}
\usepackage{amssymb}
 \usepackage{subfigure}
 \usepackage{epsfig}
 \usepackage{epstopdf} 
\usepackage{amsmath,amsfonts}
\usepackage{graphicx}
\begin{document}
	\title{\Large Effect of nonfactorizable background geometry on thermodynamics of clustering of galaxies 
}
	\author {Abdul W. Khanday}
	\email{abdulwakeelkhanday@gmail.com}
	\affiliation{{Department of Physics, National Institute of Technology  Srinagar, Jammu and Kashmir -190006, India.}}
	\author {Sudhaker Upadhyay}
	\email{sudhakerupadhyay@gmail.com}
	
	\affiliation{Department of Physics, K. L. S. College, Nawada, Bihar 805110, India}
\affiliation{Department of Physics, Magadh University, Bodh Gaya,
 Bihar  824234, India}
	\affiliation{Inter-University Centre for Astronomy and Astrophysics (IUCAA) Pune, Maharashtra-411007 }
 \affiliation{School of Physics, Damghan University, Damghan, 3671641167, Iran}
	
	\author { Hilal A. Bagat}
	\email{drhilalphst07@gmail.com}
	\affiliation{{Department of Physics, National Institute of Technology  Srinagar, Jammu and Kashmir -190006, India.}}

		\author {Prince A. Ganai}
	\email{princeganai@nitsri.net}
\affiliation{{Department of Physics, National Institute of Technology  Srinagar, 
Jammu and Kashmir -190006, India.}}

\begin{abstract}
We study the effect of nonfactorizable background geometry on the thermodynamics of the clustering of 
galaxies. A canonical partition function is derived for the gravitating system of 
galaxies treated as
point particles contained in cells of appropriate dimensions. Various thermodynamic 
equations of
state, like Helmholtz free energy and entropy, among others, are also obtained. We 
also estimate the
effect of the corrected Newton's law on the distribution function of galaxies. 
Remarkably,
the effect of the modified Newton's law is seen only in the clustering parameter 
while the standard
structure of the equations is preserved. A comparison of the modified clustering 
parameter ($b^*$) with that of the original clustering parameter is made to   
visualize the effect of the correction on the time scale of clustering. The 
possibility of system symmetry breaking is also analyzed by investigating the 
behavior of the specific heat with increasing system temperature.
	\end{abstract}	
		\maketitle
\section{Introduction}
The clustering tendency of nebulae has been studied right from the 18th century with 
the cataloging of the observed objects by C. Messier and W. Herschel. At that time, 
the question that needed an answer was whether the nebulae were internal or external 
to the Milky Way. The answer came with  
Hubble's  work ~\cite{1}, wherein he proved that the concentrations seen were indeed 
systems like the Milky Way in their own right. Once the extra-galactic nature of 
nebulae was established, the galaxies were treated as
physical systems by astronomers. In 1927,  Lundmark ~\cite{2} started the 
investigation on how the clusters form.  Zwicky ~\cite{3} was the first to estimate 
the mass of a cluster and proposed the mass discrepancy
found in the clusters. In his work, Zwicky found that the mass should be greater by 
factors of 200--400 than the visible mass. He proposed that there is an invisible 
mass, which he called ``dark matter", in the clusters that has gravitational 
attraction but is otherwise non-detectable. Spitzer and Baade ~\cite{4} proposed the 
collisional stripping theory, which leads to the study of clustering of galaxies as 
laboratories. These days,   study of clustering of galaxies is very important astronomical 
laboratories on the largest scale, with a well characterized
physical environment.

Multi-wavelength observational studies of galaxy clusters has provided tremendous 
information about the processes going on in the core of galaxy clusters. In recent 
past most of the information about the
galaxy clusters has been obtained through X-ray spectroscopy. This is due to the high 
temperature (several KeV per particle) emission of the intracluster medium. 
Along with the spectroscopic observations much theoretical study has been made to 
understand and characterize this large-scale equilibrium structure. Theoretical 
models of clusters employ various techniques focused on the understanding of 
different properties of clusters. A simple model by Kaiser ~\cite{5} which 
approximates cluster formation as dark matter driven dissipationless collapse of 
initial over-densities in an expanding universe. The predictions of this model are 
quite close to observations.  

Saslaw and   Hamilton in 1984~
\cite{6} developed a new theory for clustering in an expanding universe. This model 
is based on thermodynamics of the gravitating systems and applies to nonlinear regime 
of clustering. This model predicts the distribution function of all orders from voids 
to thousands of galaxies. The main result of this theory is the probability 
distribution of finding $N$ galaxies in a volume $V$ of any shape given by
\begin{equation}
f(N)=e^{-\bar{N}(1-b)-Nb}\frac{\bar{N}(1-b)}{N!}\left[\bar{N}(1-b)+Nb\right]^{N-1},
\end{equation} 
where $\bar{N} = nV$ is the average number of particles expected in volume $V$ with 
average density $n$. The
constant $b$ is the correlation parameter and can take values between $0$ and $1$. The applicability of thermodynamics to cosmological many-body problems paves way for the applicability of statistical mechanics.
In this procedure analytical expressions for the partition function (grand canonical) is derived for the
galaxies treated as point particles as well as for galaxies-with-halos (extended masses)~\cite{7}.
Recently, many modifications to general theory of relativity have been put forward~\cite{8}. These models
have implications for the evolution of density fluctuations in the early universe that has caused the large
scale cosmic structure. For instance, $f(R)$ theory is one such candidate in which an additional term, a
function $f(R)$ of Ricci curvature is added to Einstein-Hilbert action ~\cite{9}. These models give enhanced
gravitational force on scales relevant to structure formation and hence an enhanced structure formation
on these scales. There has also been progress in studying the effect of modified Newtonian potential on
the clustering of galaxies \cite{10,10l,11,11a,11b, 11c, 10a,10b,10c}.
\par
Conventionally we believe that Newton's force law for gravity implies only four non-compact dimensions. This is also verified by the fact that the standard model matter cannot propagate in extra
dimensions to a large distance without contradicting with the observations. This problem can be avoided
if we confine standard model in a ``$3$-brane” i.e. $3+1$-dimensional subspace. However, in case of gravity this model is not possible as gravity, being dynamics of spacetime itself, propagates in all dimensions. If we apply the above framework to gravity, the size of the extra dimensions should be sufficiently small (compactification) so that the model agrees with the current gravitational tests. These properties are a consequence of the fact that the metric of the four non-compact dimensions is independent of the coordinates in extra dimensions i.e. factorizable geometry. If we drop this assumption and take $4+n$ non-compact dimensions with a non-factorizable background geometry, the scenario changes significantly.  Randall and Sundrum ~\cite{12} in their work took $n=1$ ($u$-direction) i.e. $4+1$ non-compact dimensions and successfully reproduced an effective four dimensional theory of gravity with the potential
given by
\begin{equation}
	\phi(r)=G\frac{m^2}{r}\left(1+\frac{1}{r^2k^2}\right).\label{02}
\end{equation}
The leading term in equation (\ref{02}) is the usual Newtonian potential and the second term is the correction generated due to Kaluza-Klein modes. This form of potential comes from the  Randall–Sundrum geometry
\begin{eqnarray}
ds^2=a^2(u)\eta_{\mu\nu}dx^\mu dx^\nu-du^2,
\end{eqnarray}
where $\eta_{\mu\nu}$ is the four-dimensional Minkowski metric and $a (u)=e^{-k|u|}$ is the warp factor. This metric  is nonfactorizable as
 this, unlike the usual Kaluza-Klein scenarios, can not be described by  product of the four-dimensional Minkowski
space and a (compact) manifold of extra dimensions.

 Although  galactic clustering under modified theories of gravity has been studied extensively, the study under the effect of  nonfactorizable background geometry is not yet  studied. This provides us an opportunity to bridge this gap. This is a motivation of present study. 
We present our study in the following manner. In section II, we construct the partition function for the system of galaxies. In section III, we quantitatively study the effect of the extra term ($ r$ dependency)  in  Newton's law  on various thermodynamic equations of state, viz. free energy, entropy, and chemical potential, etc. We study the effects of this force form   on  the  statistical distribution of   galaxies in section IV. The behavior of specific heat as an indicator of  possible phase transition is also studied in section V. The effect of the correction term yields a modified clustering parameter which estimates the extent of correlation among system particles. Finally we discuss the summary and future prospectus in the last section. 

\section{GENERALIZED PARTITION FUNCTION}
Here, in this section, we deduce the partition function of the system of galaxies treated as homogeneous over large regions in an expanding universe as has been previously done in Ref. ~\cite{7}. Our system consists of larger number of cells (ensemble of cells) with same volume $V$ or radius $R$ ($R<<V$), and have average number density
$\bar{N}$. Also, we let the particle number and their total energy vary among the cells so that it represents the grand canonical ensemble. In this system the galaxies (particles) will interact only pairwise through gravitational force and over a large space or region the distribution of particles (galaxies) is statistically homogeneous.
\par
The general form of the partition function of a system of $N$   galaxies  of equal mass $m$ interacting gravitationally with a potential energy $\phi$  having average temperature $T$ and momenta $ p_{i}$ is given by
\begin{equation}  	
Z_{N}(T,V)=
\frac{1}{\lambda^{3N}N!} \int \exp{\left[-T^{-1}\left( \sum_{i=1}^{N}\frac{p_{i}^{2}}{2m}+\phi(r_{1},r_{2},...r_{N})\right)\right]} d^{3N}p d^{3N}r.\label{03}
\end{equation}
Here $N!$ takes care of distinguishability of classical system of particles and $\lambda$ is a normalization factor.  Boltzmann's constant is set unit here.  Integrating over the momenta space, equation (\ref{03}) simplifies to,
\begin{equation}
\hspace{2cm} Z_{N}(T,V)=\frac{1}{N!}\left(\frac{2\pi mT}{\lambda^2}\right)^\frac{3N}{2} Q_{N}(T,V),
\end{equation}
	where configurational integral
\begin{equation}
 Q_{N}(T,V)=\displaystyle \int...\int  \exp  \left[-T^{-1}\phi(r_{1},r_{2},...r_{N})\right] d^{3N}r.\label{5}
\end{equation}
  Generally, the gravitational potential energy
function $\phi(r_{1},r_{2},...r_{N})$ is a function of the relative position vector $r_{ij}=|r_{i}-r_{j}|$ and is the sum of
potential energy of all the pairs of particles. Here our main task is to evaluate the integral  $Q_{N}(T,V)$.
In our system of gravitating bodies the potential energy $\phi(r_{1},r_{2},...r_{N})$ of the system is due to all pairs of particles (galaxies) of which the system is made, i.e.
\begin{equation}
\phi(r_{1},r_{2},...r_{N})=\sum_{1\le i\le j\le N}^{}\phi(r_{i,j})=\sum_{1\le i\le j\le N}^{}\phi_{ij}(r).
\end{equation}
With this simplification, equation (\ref{5}) can now be written as 
 \begin{equation}
\hspace{2cm} Q_{N}(T,V)=\displaystyle\int...\int\displaystyle \prod_{1\le i\le j\le N}^{}\left[-T^{-1}\phi_{ij}(r)\right] d^{3N}r,\label{7}
\end{equation}
where $\phi_{ij}$ represents the potential energy due to gravitational interaction between the $i^{th}$ and $j^{th}$ particle. In order to solve the confugirational integral, we make use of the usual two-particle
function defined as 
\begin{equation}
f_{ij}=e^{ -\frac{\phi_{ij}}{T}}-1.
\end{equation}
The function $f_{ij}$ is identically zero if there is no interaction between the particles (galaxies) e,g ideal gases, and is non zero in presence of interaction. Also at extremely high temperature it is extremely
small comparison with unity. With the substitution of two point function, $f_{ij}$, equation (\ref{7}) takes the form 
\begin{equation}
Q_{N}(T,V)=\int...\int\prod_{1\le i\le j\le N}^{}(1+f_{ij})d^{3}r_{1}d^{3}r_{2}...d^{3}r_{N}.\label{9}
\end{equation}
In the above integral, the higher order terms like $\sum f_{ij}  f_i^{\prime}j^{\prime}$ can be dropped as these represent the
interaction of more than two particles at once. However, in gravitating system all particles interact pairwise only, therefore the product in the equation (\ref{9}) can be represented as
 \begin{equation}
\displaystyle \prod_{1\le i\le j\le N}(1+f_{ij})=\displaystyle \prod_{1\le i} \prod_{i< j\le N}^{}{(1+f_{ij})} =\displaystyle\prod_{j=1,2,3,4...N}^{}(1+f_{1j})(1+f_{2j})(1+f_{3j})...(1+f_{Nj}).
\end{equation}
This sum excludes or neglects the terms involving self energy like $f_{jj}$. Hence when $j=2$ we have only one term $(1+f_{12})$. For  $j=3$, we have just two terms $(1+f_{12})(1+f_{23})$ and so on.Therefore for other values of $j$ the above equation becomes
 \begin{equation}
Q_{N}(T,V)=\displaystyle\int...\int(1+f_{12})(1+f_{13})(1+f_{23})(1+f_{14})...(1+f_{N-1,N})d^{3}r_{1}d^{3}r_{2}...d^{3}r_{N}.\label{11}
\end{equation}
Now, the main task remains the evaluation of this integral  for various values of $N$ and then
generalize it. 

For point masses the particle function diverges when it includes energy states corresponding
to $r_{ij}=0$, which in turn results in the divergence of Hamiltonian of the system. In order to remove this
divergence we introduce a new parameter $\epsilon$, this $\epsilon$ is called the softening parameter. The typical value of
softening parameter $\epsilon$ is $0.01\le \epsilon \le 0.05$ in units of the constant cell size~\cite{7}.
 With the incorporation of softening parameter the modified Newtonian potential (\ref{02}) can be written as
\begin{equation}
\phi_{ij}= Gm^{2}\left[\frac{1}{(r^2_{ij}+\epsilon^{2})^{1/2}}+\frac{1}{(r^6_{ij}+\epsilon^{6})^{1/2}k^2}\right].
\end{equation}
Using this value of $\phi_{ij}$,  we get the following two-particle function:
\begin{equation}
f_{ij}+1=e^{ \frac{  Gm^{2}}{(r^2_{ij}+\epsilon^{2})^{1/2}T}+\frac{ Gm^{2}}{(r^6_{ij}+\epsilon^{6})^{1/2}Tk^2} }.
\end{equation}
This can further be expanded to get,
\begin{equation}
\large	f_{ij}=\frac{Gm^{2}}{T}\left[\frac{1}{(r^2_{ij}+\epsilon^{2})^{1/2}}+\frac{1}{(r^6_{ij}+\epsilon^{6})^{1/2}k^2}\right].\label{14}
\end{equation}
Since the system is not virialized on all scales, so the above expansion is dominating  up to linear order only.
Using equation (\ref{14}) in equation (\ref{11}), values of $Q_N$ for different values  of $N$ can be obtained. For $N = 1$, we have
\begin{equation}
Q_{1}(T,V)= V.
\end{equation}
For $N=2$,   first we fix the position of $r_1$ and evaluate the above integral over all the other particles. The integral simplifies to 
 \begin{equation}
Q_{2}(T,V)=V^{2}\bigg[1+Y (\beta_{1}+\beta_{2} )\bigg],
\end{equation}
where

\hspace{2cm}$\beta_{1}=\sqrt{1+\frac{\epsilon^2}{R_1^2}}+
\frac{\epsilon^2}{R_1^2} \log\frac{\epsilon/R_1}{1+\sqrt{1+\epsilon^2/
R_1^2}},$

\hspace{2cm}$\beta_{2}=\frac{1}{3R_1^2 k^2}\log\bigg[\frac{1+\sqrt{1+
\epsilon^6/R_1^6}}{\epsilon^6/R_1^6}\bigg],$

\hspace{2cm}$ Y=\frac{3Gm^{2}}{2TR_{1}}.$

Proceeding in the same way, the value of $Q_N$ for $N = 3, 4, 5, ..., N$ can be obtained. For $N = 3$, we get
\begin{equation}
Q_{3}(T,V)=V^{3}\bigg[1+Y (\beta_{1}+\beta_{2} )\bigg]^{2}.
\end{equation}
Generalizing above equation for $N$ particles, we obtain
\begin{equation}
Q_{N}(T,V)=V^{N}\bigg[1+Y (\beta_{1}+\beta_{2} )\bigg]^{N-1}.\label{19}
\end{equation}
Finally, substituting equation (\ref{19}) into equation (\ref{03}), we get the partition function for gravitating system of $N$ particles (galaxies) under the modified Newtonian potential as
\begin{equation}
Z_{N}(T,V)=\frac{1}{N!}\bigg(\frac{2\pi mT}{\Lambda^{2}}\bigg)^{\frac{3N}{2}} V^{N}\bigg[1+ (\beta_{1}+\beta_{2} )Y\bigg]^{N-1}.\label{20}
\end{equation}
This is the standard form of canonical partition function for a system of $N$ particles interacting through the modified Newton's law. The modification due to the incorporation of nonfactorizable background geometry is inherent in the parameter $\beta_{2}$. 

\section{ THERMODYNAMIC EQUATIONS OF STATE}
\begin{figure}[h!]
	\centering
	\includegraphics[width=8 cm, height=6 cm]{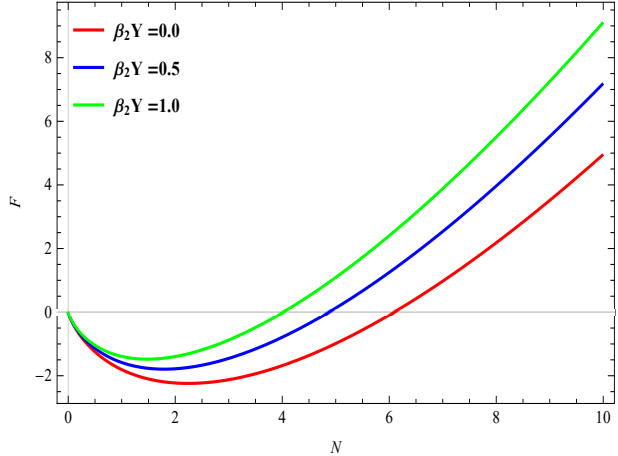}
	\caption{Variation of free energy with particle number for different values of the correction term $\beta_{2}Y$.}\label{fig1}
\end{figure}

Once the partition function is known, it a matter of calculation to deduce various thermodynamic quantities. For example, the Helmholtz free energy for our system of galaxies can be obtained using the general statistical relation $F=-T\ln Z_{N}(T,V)$ along with equation (\ref{20}). The free energy for the system of galaxies takes the following form: 
\begin{equation}
F=NT\ln\bigg(\frac{N}{V}T^{\frac{-3}{2}}\bigg)-NT-\frac{3}{2}NT\ln\bigg(\frac{2\pi mT}{\lambda^{2}}\bigg)-NT\ln [1+Y(\beta_{1}+\beta_{2})].
\label{21}
\end{equation}
Here, approximation $N-1\approx N$ is used. The effect of the correction term can also be depicted from the  Fig. \ref{fig1}. Here, we see that correction term
increases the value of free energy. 

 For a given free energy, we can study the other  thermodynamic equations of state easily. The entropy of the system can be obtained using the fundamental relation, $S=-\bigg(\frac{\partial F}{\partial T}\bigg)_{V,N}$. For the Helmholtz free energy (\ref{21}), the entropy of the system is calculated as
\begin{equation}
S=N\ln\biggl(\frac{V}{N}T^{3/2}\biggl) +N \ln [1+Y (\beta_{1}+\beta_{2} ) ] -3N\frac{Y (\beta_{1}+\beta_{2} )}{ 1+Y (\beta_{1}+\beta_{2} ) } + \frac{5}{2}N +\frac{3}{2}N \ln\bigg(\frac{2\pi m }{\lambda^{2}}\bigg).\label{22}
\end{equation}
\par
\begin{figure}[h!]
	\centering
	\includegraphics[width=8 cm, height=6 cm]{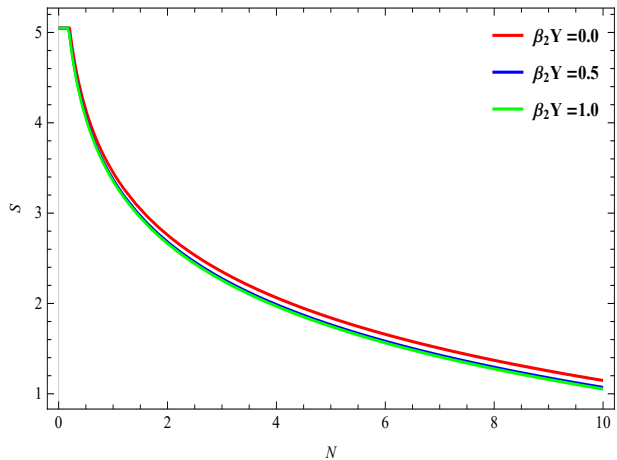}
	\caption{ The graph shows the variation of entropy as a function of particle number for different values of correction parameter $\beta_2Y$.}\label{fig3}
\end{figure}
\begin{figure}[h!]
	\centering
	\includegraphics[width=8 cm, height=6 cm]{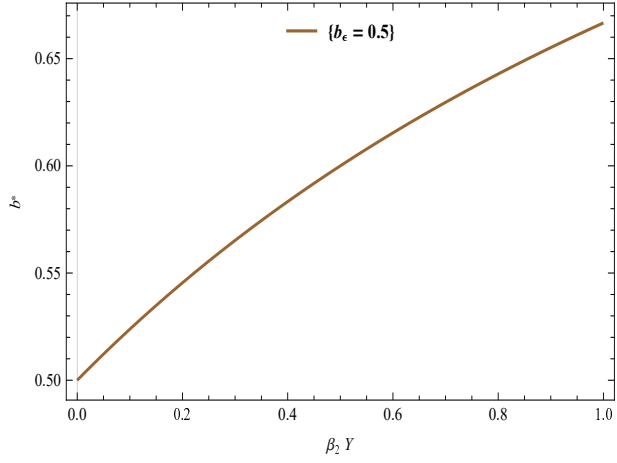}
	\caption{ The  variation of the clustering parameter $b^*$ with an increase in the strength of the correction term  $\beta_2Y$ for a fixed value of the unmodified clustering parameter $b_\epsilon$.}\label{fig2}
\end{figure}

The effect of the correction parameter $b^*Y$ on the entropy of the system of galaxies can be depicted from the graph (Fig. \ref{fig3}). 
This equation   can further be  simplified to entropy per particle as
\begin{equation}
\frac{S}{N}=\ln\biggl(\frac{V}{N}T^{3/2}\biggl)-\ln[1-b^*]-3b^*
+\frac{5}{2}+\frac{3}{2}\ln \frac{2\pi m}{\lambda^2},
\end{equation}
where \begin{equation}
b^*=\frac{\left(\beta_1+\beta_2\right)Y}{ 1+\left(\beta_1+\beta_2\right)Y },\label{b}
\end{equation} 
is the modified clustering parameter that takes values between $0$ and $1$ and 
estimates the strength of correlation between system particles. The standard 
clustering parameter $b_\epsilon$, defined  as $b_\epsilon=\frac{\beta_1Y}{1+\beta_1Y}$   ~\cite{7}, is a limiting case of the 
  modified parameter $b^*$ given in (\ref{b}). These parameters are related as
\begin{equation}
b^*= \frac{b_\epsilon (1-\beta_2Y)+\beta_2Y}{1+\beta_2Y-b_\epsilon \beta_2Y}.
\end{equation} 
 From the above relation, it is evident that $b^*\rightarrow b_\epsilon$ when $\beta_2\rightarrow 0$.
 The variation of the  modified clustering parameter $b^*$ with the strength of the correction term can be visualized from the figure \ref{fig2}. The clustering becomes  stronger as the value of the correction term increases. This, in turn, can  affect the time-scale of clustering.   
 
In order to study the internal energy of the system of galaxies, we use the basic definition of internal energy, $ U = F + T S$. After substituting the  calculated values of free energy (\ref{21})  and entropy  (\ref{22}), this results
\begin{eqnarray}
U&=&\frac{3}{2}NT\bigg[1-2\frac{Y (\beta_{1}+\beta_{2} )}{1+Y (\beta_{1}+\beta_{2} )}\bigg]\nonumber\\
&=&\frac{3}{2}NT\left[1-2b^*\right].\label{26}
\end{eqnarray}
 \begin{figure}[h!]
	\centering
	\includegraphics[width=8 cm, height=6 cm]{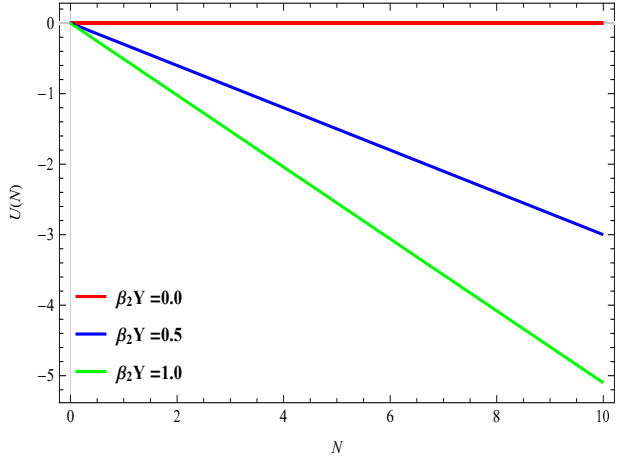}
	\caption{ The  variation of internal energy $U$ as a function of particle number for different values of correction parameter $\beta_2Y$.}\label{fig4}
\end{figure}
The graphical representation of the effect of the correction parameter on the internal energy of  the system of galaxies can be seen in Fig. \ref{fig4}. 

\begin{figure}[h!]
	\centering
	\includegraphics[width=8 cm, height=6 cm]{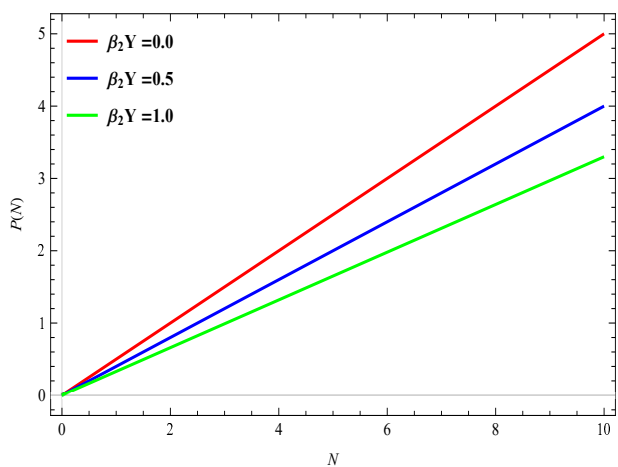}
	\caption{ The  variation of entropy as a function of particle number for different values of correction parameter $\beta_2Y$.}\label{fig5}
\end{figure}

The pressure of the system can be calculated, using the fundamental definitions $P=-\bigg(\frac{\partial F}{\partial V}\bigg)_{T,N}$. 
 Here,  pressure is calculated as
\begin{align}
P&=\frac{NT}{V} \left[1-\frac{Y (\beta_{1}+\beta_{2} )}{1+Y (\beta_{1}+\beta_{2} )} \right],
\nonumber\\ 
&=\frac{NT}{V}\left[1-b^*\right].\label{27}
\end{align}
The behavior of pressure of system with increasing particle number for different values of correction parameter $b^*Y$ can be visualized from the  Fig. \ref{fig5}.
Here, we clearly see that correction term decreases the entropy of the system.
\par 
The chemical potential of the system can be calculated using the relation 
 $\mu=\bigg(\frac{\partial F}{\partial N}\bigg)_{T,V}$ as
\begin{eqnarray}
\mu &=&T\bigg(\ln\frac{N}{V}T^{-\frac{3}{2}}\bigg)+T\ln\bigg[1-\frac{Y (\beta_{1}+\beta_{2} )}{1+Y (\beta_{1}+\beta_{2} )}\bigg]-T\frac{Y (\beta_{1}+\beta_{2} )}{1+Y 
(\beta_{1}+\beta_{2} )}-\frac{3}{2}T\ln\bigg(\frac{2\pi m}{\lambda^{2}}\bigg),\nonumber\\
&=&T\left(\ln \frac{N}{V}T^{-3/2}\right)+T\ln \left[1-b^*\right]-Tb^*
-\frac{3}{2}T\ln\left(\frac{2\pi M}{\lambda^2}\right).\label{29}
\end{eqnarray} 
\begin{figure}[h!]
	\centering
	\includegraphics[width=8 cm, height=6 cm]{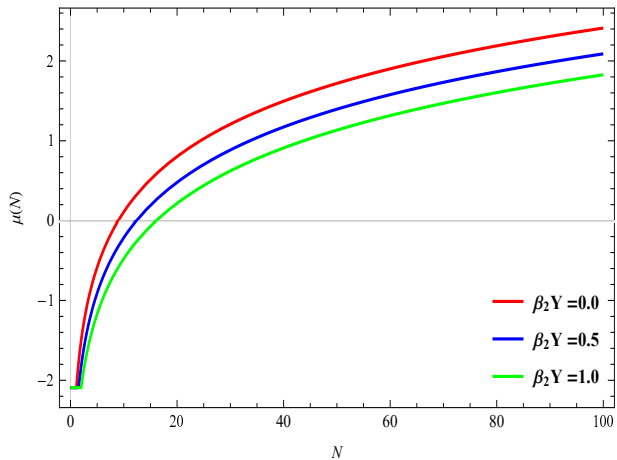}
	\caption{ The graph shows the variation of chemical potential $\mu$ as a function of particle number for different values of correction parameter $\beta_2Y$.}\label{fig6}
\end{figure}
The figure \ref{fig6} shows  the graphical variation of the chemical potential $\mu$ with increasing particle number for different values of correction parameter $b^*Y$.
The equations of state given in  Eqs. (\ref{22}), (\ref{26}), (\ref{27}) and (\ref{29}) contain modified clustering parameter $b^*$ instead of $b_\epsilon$ as used in the original expressions~\cite{7}.  
\section{General distribution function}
\begin{figure}[h!]
	\centering
	\includegraphics[width=8 cm, height=6 cm]{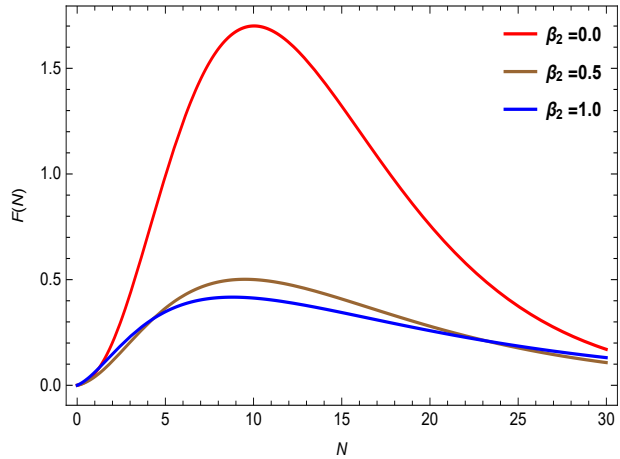}
	\caption{ Behavior of the distribution function $F(N)$ as a function of particle number, $N$.   Red curve corresponds to $\beta_{2}=0$, i.e., no correction. 	Brown line corresponds to $\beta_{2}=0.5$. Blue line corresponds to $\beta_{2}=1$.   }
	\label{fig7}\end{figure}

The probability distribution function $f(N)$ characterizes the galactic clustering as it contains the voids distribution as well as the number of particles (galaxies) in cells of a given size distributed through out
the system. The grand partition function is defined by  
\begin{equation}
	Z_G(T,V,z)=\sum_{N=0}^{\infty}\exp \left(\frac{N\mu}{T}Z_N(T,V)\right).
\end{equation} 
 The distribution function   of finding $N$ particles in a volume of a cell $V$ of grand canonical ensemble is given by
\begin{eqnarray}
F(N)&=&\sum_{i=0}^{N}\frac{\exp\frac{N\mu}{T}\exp\frac{-U}{T}}{Z_G(T,V,z)},\nonumber\\
&=&\frac{\exp\frac{N\mu}{T}Z_N(T,V)}{Z_G(T,V,z)}.\label{32}
\end{eqnarray} 
Here, the weight factor $z=\exp\frac{\mu}{T}$ is the activity that represents the average value of $\bar{N}$. From this basic relation, we can calculate the distribution function. Exploiting the partition function (\ref{20}) and the chemical potential (\ref{29}), (\ref{32})  the distribution function for the system of galaxies is given as
\begin{equation}
F(N)= \frac{\bar{N}}{N!}(1-b^*)\bigg[\bar{N}(1-b^*)+Nb^*\bigg]^{N-1} \exp-Nb^*-\bar{N}(1-b^*).
\end{equation}
The basic structure of the distribution function is same as derived by Saslaw and Hamilton in Refs. \cite{7, 13}. The behavior of the distribution function as a function of particle number $N$ in two-dimensions  is shown in figure \ref{fig7}. From the figure, it can be seen that the peak of the distribution function has flattened as the value of the correction parameter   increases.  We also infer that the correction parameter $\beta_{2}$ shifts the peak downwards without changing the basic structure of the curve.
\section{The behavior of specific heat as an indicator of phase transition}
The Poisson distribution  (zero correlation) of a many-body system, driven gravitationally, evolves through many stages from zero correlation to some positive value of the correlation parameter $b^*$. This evolution can be characterized as a form of phase transition from uncorrelated phase to correlated phase, i.e., $b^*=0$ to $b^*>0 $. Through this phase transition the homogeneity of the system is lost and lumps of particles are created.  
 As an important indicator of phase transition, we analyze the variation of specific heat with temperature, $T$, of the system.  

\begin{figure}[h!]
	\centering
	\includegraphics[width=8 cm, height=6 cm]{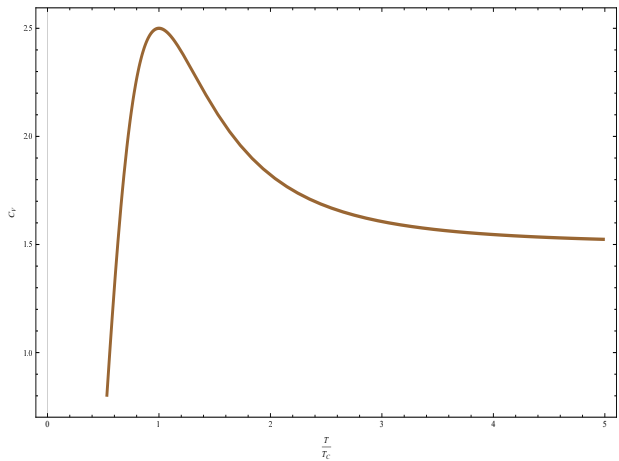}
	\caption{The variation of specific heat $C_V$ with  $T/T_C$. The system symmetry breaks around the critical temperature ($T=T_C$). }\label{fig8}
\end{figure}

 The specific heat (at constant volume) $C_V$ is defined as
\begin{equation}
	C_V=\frac{1}{N}\left(\frac{\partial U}{\partial T}\right)_{V,N}. 
\end{equation}
For internal energy (\ref{26}), the above equation leads to the specific heat of the system as
\begin{equation}
	C_V=\frac{3}{2}\left[\frac{1+6\beta Y -4\beta^2Y^2}{\left(1+\beta Y\right)^2}\right],\label{35}
\end{equation} 
where $\beta=\beta_{1}+\beta_{2}$.
At $b^*=0$, the specific heat takes the value $C_V=3/2$. This corresponds to zero correlation of the system particles. At $b^*=1$, $C_V=-3/2$, which corresponds to fully virialized system. In between these two extreme values of the correlation parameter $b^*$, the specific heat takes an extreme value which corresponds to a phase transition within the system corresponding to a critical value of temperature, $T=T_C$, i.e.,
\begin{equation*}
	\frac{\partial C_V}{\partial T}\biggr\rvert_{T=T_C}=0.
\end{equation*} 
 In the modified potential the correlation parameter takes the value
\begin{equation}
	T_C=\left[3\frac{\bar{N}}{V}\left(GM^2\right)^3\left(\beta_{1}+\beta_{2}\right)\right]^{1/3}.
\end{equation}
The specific heat $C_V$  given in (\ref{35})  can be expressed in terms of the above critical temperature as 
\begin{equation}
 C_V=\frac{3}{2}\left[1-2\frac{1-4\left(T/T_C\right)^3}{\left\{1+2\left(T/T_C\right)^3\right\}^2}\right].
\end{equation}
At $T=T_C$, the specific heat becomes $C_V=5/2$, which is a property of a diatomic gas. This corresponds to the formation of binary systems and indicates the system symmetry breaking at average inter-particle separation. At this point the hierarchical phase transition occurs at the lowest scale and propagates to higher scales in the system. This behavior of specific heat is also evident from the  Fig. \ref{fig8} in which the specific heat varies finitely around critical temperature.    
\section{Conclusion}
In this paper, we have analyzed the clustering of galaxies under a modified Newtonian potential motivated by the inclusion of nonfactorizable background geometry. Here,
we have considered a strongly interacting system of galaxies and derived the gravitational partition
function for galaxies interacting under the modified gravitational potential. We have also computed
various thermodynamic equations of state for the system of galaxies. A general clustering parameter  was also obtained for the system of galaxies. We observed that the correlation parameter gets stronger and stronger as the strength of the correction factor is increased. This behavior of the clustering parameter   can affect the time scale of clustering of the galaxy systems.  The effect of modification in the potential has also affected the distribution function of the system of galaxies.
The corrected distribution function has a low peak as compared to the original distribution function.  
The evolution of the system with an increasing system temperature is characterized by the behavior specific heat implying a possibility of  phase transition around the critical temperature.

  \end{document}